\documentclass[prl,aps,preprint,showpacs]{revtex4}
\usepackage{latexsym} 

\begin{document}

\title
{Covariance of the Schr\"odinger equation under low velocity boosts.}

\author
{A. B. van Oosten}

\affiliation{Theoretical Chemistry and Materials Science Centre, University of Groningen, Nijenborgh 4, Groningen 9747 AG, The Netherlands}

\date{\today}

\begin{abstract}
It is known that the Schr\"odinger equation is not covariant under Galilei boosts,
unless the phase of its solutions are shifted simultaneously.
It is argued that the phase shift is not a coordinate transformation,
because it depends on the mass of the Schr\"odinger particle.
The phase shift also cannot be derived from low speed Lorentz boost.
It is proposed to extend the Galilei boost with two terms of order $\frac{v^2}{c^2}$
to avoid these issues and to guarantee covariance of the Schr\"odinger kinetic energy and  momentum.
The extensions imply that proper time and relativity of simultaneity are essential features of Schr\"odinger quantum mechanics.
\end{abstract}


\maketitle

The Schr\"odinger equation \cite{hbar},
\begin{equation}
-i \partial_t \psi (t,\vec{r}) = - \frac{\nabla^2}{2m} \psi (t,\vec{r}),
\label{schrod}
\end{equation}
is not covariant under Galilei boosts.
Consider a solution describing a free massive neutral particle,
\begin{equation}
\psi (t,\vec{r}) = e^{i \frac{p^2}{2m} t - i \vec{p} \cdot \vec{r}}.
\label{psi}
\end{equation}
A Galilei boost,
 \begin{eqnarray}
t' &=& t , \nonumber \\
\vec{r} \,' &=& \vec{r} -\vec{v} t ,
\label{Gboost}
\end{eqnarray}
transforms this into
\begin{eqnarray}
\psi (t',\vec{r} \, ') = e^{i \vec{p} \cdot \vec{v} t } \psi (t,\vec{r}),
\label{psi-Gb}
\end{eqnarray}
which has incorrect energy and momentum.
An equivalent statement of the problem can be made using the fact
 that the Schr\"odinger energy and momentum operators 
$-i \partial_t $ and $-i \vec{\nabla}$ in Galilei relativity transform as
\begin{eqnarray}
{\partial'}_t &=& \partial_t + \vec{v} \cdot \vec {\nabla} , \nonumber \\
\vec{\nabla} \, ' &=& \vec{\nabla}.
\label{Gboost-cov}
\end{eqnarray}
According to the correspondence principle energy and momentum transform follow the same transformation law in classical non-relativistic and in Schroedinger
mechanics. Eq. (\ref{Gboost-cov}) is at variance with this.

The known solution \cite{Brown} to this problem is
to extend the Galilei transformation with a phase shift
and redefine the wave function in a new frame as
\begin{equation}
\psi' (t',\vec{r} \, ')
= e^{i ( \frac{1}{2}m v^2  t - m \vec{v} \cdot \vec{r} )}
\psi (t,\vec{r}-\vec{v}t).
\label{mod-boost}
\end{equation}
which leads to the correct result
\begin{equation}
\psi ' (t',\vec{r} \, ')
= e^{i \frac{(p+m\vec{v})^2}{2m} t - i (\vec{p} +m\vec{v})\cdot \vec{r}}
\psi (t,\vec{r}) .
\label{psi-boost}
\end{equation}

Greenberger \cite{Green} has shown that effects of proper time persist 
in Schr\"odinger quantum mechanics. 
In his example two identical systems, each described by a superposition of two Schr\"odinger eigenfunctions of different energy, undergo different accelerations. 
This leads to an in principle observable phase difference.

In this paper I argue that the phase shift approach is not acceptable
and that proper time effects are more fundamental even than proposed before \cite{Green}.
Proper time as well as relativity of simultaneity needs to be taken taken into account 
in low velocity boosts
in order to obtain covariance of the Schr\"odinger equation.

The phase shift approach can be criticised on the following two grounds.
Firstly, the transformation (\ref{mod-boost}) depends on the mass of the Schr\"odinger particle,
so that explicitly different transformations are needed for particles of different mass.
Therefore Eqs. \ref{mod-boost} do not constitute a coordinate transformation, 
Secondly, the phase shift operation does not follow 
from a Lorentz transformation for low velocity.
This illustrates its ad hoc character. 

The derivation of a suitable coordinate transformation that describes a low velocity boost
and leads to the correct boosted energy and momentum proceeds in two steps.
Firstly, the wave function (\ref{psi}) can not be regarded 
as the low velocity limit of a relativistic wave function.
This can be repaired by including the rest energy into Eq. (\ref{schrod}), which becomes
\begin{equation}
-i \partial_t \psi (t,\vec{r}) = ( - \frac{\nabla^2}{2m} + mc^2) \psi (t,\vec{r}) ,
\label{schrod-m}
\end{equation}
The equivalent of (\ref{psi}) then becomes
\begin{equation}
\psi (t,\vec{r}) = e^{i mc^2t + i \frac{p^2}{2m} t - i \vec{p} \cdot \vec{r}}.
\label{psi-m}
\end{equation}
Note that this equation, too, is not covariant under Eqs. (\ref{Gboost}). 
Secondly, it is noted that in an expansion of the Lorentz boost terms of order $c^{-2}$ 
contribute terms of order $c^0$ in the final result if they multiply with the rest frame energy.
Therefore these must be retained in the transformation, which leads to
\begin{eqnarray}
t' &=& (1 + \frac{1}{2} \frac{v^2}{c^2}) t -  \frac{\vec{v} \cdot \vec{r}}{c^2} , \nonumber \\
\vec{r} \, ' &=& \vec{r} -\vec{v} t .
\label{boost}
\end{eqnarray}
Further terms of order $\frac{v^2}{c^2}$ do not result in low-velocity effects 
since they are involved in a product with the rest energy.
When this is applied to the wavefunction (\ref{psi-m}), one obtains the result
\begin{equation}
\psi (t',\vec{r}') = e^{ i mc^2t + i \frac{(\vec{p}-m\vec{v})^2}{2m} t - i (\vec{p} - m \vec{v}) \cdot \vec{r} }.
\label{psi-m-boost}
\end{equation}
Eqs. (\ref{boost}) can be expressed in matrix form as
\begin{equation}
T(\vec{v}) =  \left( \begin{array}{cccc}
                      1 + \frac{1}{2} \frac{v^2}{c^2} & - \frac{v_x}{c^2} & - \frac{v_y}{c^2}  & - \frac{v_z}{c^2} \\
                       - v_x & 1 & 0 & 0 \\
                       - v_y & 0 & 1 & 0 \\
                       - v_z & 0 & 0 & 1
                       \end{array} \right)
\end{equation}
The inverse boost is $T(-\vec{v})$ as can be seen from
\begin{equation}
T(-\vec{v}) T(\vec{v}) =  
\left( \begin{array}{cccc}
1 + O (\frac{v^4}{c^4}) & O(\frac{v^2 v_x}{c^4}) & O(\frac{v^2 v_y}{c^4}) & O(\frac{v^2 v_z}{c^4}) \\
                       O(\frac{v^2 v_x}{c^2}) & 1+O(\frac{v_x^2}{c^2}) & 0 & 0 \\
                       O(\frac{v^2 v_y}{c^2}) & 0 & 1+O(\frac{v_y^2}{c^2}) & 0 \\
                       O(\frac{v^2 v_z}{c^2}) & 0 & 0 & 1+O(\frac{v_z^2}{c^2})
\end{array} \right) ,
\end{equation}
which indeed leaves Expr. (\ref{psi-m}) invariant up to terms vanishing at least as $c^-2$.
Eqs. (\ref{boost}) imply
\begin{eqnarray}
i \partial_t' &=& (1 + \frac{1}{2} \frac{v^2}{c^2}) \partial_t +  \frac{\vec{v} \cdot \vec{\nabla}}{c^2} , \nonumber \\
\vec{\nabla} \, ' &=& \vec{\nabla} + \vec{v} \partial_t ,
\label{boost-cov}
\end{eqnarray}
to the same approximation, 
which is equivalent to Eqs. (\ref{boost}).
Thus Eqs. (\ref{boost}) satisfy the correspondence principle.

The two extra terms in Eqs. (\ref{boost}), as compared to Eqs. (\ref{Gboost}) 
involve essential aspects of special relativity.
The term $\frac{1}{2} \frac{v^2}{c^2} t $, which implies that proper time effects,
is required for the covariance of Schr\"odinger kinetic energy.
The term $\frac{\vec{v} \cdot \vec{r}}{c^2}$, which implies that simultaneity
of events at different positions is relative, 
is required for the covariance of Schr\"odinger momentum.

It is interesting to apply the present approach to the twin system example of Greenberger.
Consider a particle of mass m described by Eq. (\ref{psi-m})
that is observed from a non-inertial coordinate system.
In the spirit of Ref. \cite{Green}, Eqs. (8), it is assumed that the coordinate transformation is
\begin{eqnarray}
t' &=& (1 + \frac{1}{2} \frac{v^2}{c^2}) t -  \frac{\vec{v} \cdot \vec{r}}{c^2} , \nonumber \\
\vec{r} \,' &=& \vec{r} -\xi (t) .
\label{non-inertial}
\end{eqnarray}
In the case that $\xi (t) \not= 0$ only for $0<t<t_1$, 
the wavefunction at $t>t_1$ acquires a phase shift 
relative to a system with $\xi = 0$ at all times of
\begin{equation}
\phi  = \int_0^{t_1} dt \frac{1}{2} m \dot{\xi}^2 ,
\label{phase}
\end{equation}
which agrees with Greenberger's result.
A superposition of two wavefunctions of different energy is not needed here,
since the full rest energy is included.
As in Greenberger's example, two observers with a different non-inertial history
will in general observe a phase difference.

In conclusion, two extensions of the Galilei boost are proposed that guarantee covariance of the Schr\"odinger equation if the rest energy is included in it. 
The first extension guarantees energy covariance and is associated with proper time.
The second extension guarantees momentum covariance and is associated with
relativity of simultaneity.

\end{document}